\documentclass[aps,twocolumn,groupedaddress,showpacs]{revtex4}
\usepackage{graphicx}
\begin{document}

\title{Dynamical structure factors of the magnetization-plateau state in the $S=1/2$ bond-alternating spin chain with a next-nearest-neighbor interaction}

\author{Nobuyasu Haga and Sei-ichiro Suga}
\affiliation{Department of Applied Physics, Osaka University, Suita, Osaka 565-0871, Japan}
\date{\today}
\begin{abstract}
We calculate the dynamical structure factors of the magnetization-plateau state in the $S=1/2$ bond-alternating spin chain with a next-nearest-neighbor interaction. The results show characteristic behaviors depending on the next-nearest-neighbor interaction $\alpha$ and the bond-alternation $\delta$. 
We discuss the lower excited states in comparison with the exact excitation spectrums of an effective Hamiltonian. From the finite size effects, characteristics of the lowest excited states are investigated. 
The dispersionless mode of the lowest excitation appears in adequate sets of $\alpha$ and $\delta$, indicating that the lowest excitation is localized spatially and forms an isolated mode below the excitation continuum. 
We further calculate the static structure factors. 
The largest intensity is located at $q=\pi$ for small $\delta$ in fixed $\alpha$. With increasing $\delta$, the wavenumber of the largest intensity shifts towards $q=\pi/2$, taking the incommensurate value. 
\end{abstract}
\maketitle
\section{Introduction}
One-dimensional (1D) quantum spin systems have attracted great attention both theoretically and experimentally. Fascinating aspects can be seen in some characteristic systems under magnetic fields. 
Recently, it was shown numerically that there appears a plateau region on the magnetization curve at half of the saturation value in the $S=1/2$ bond-alternating spin chain with a next-nearest-neighbor (NNN) interaction~\cite{tonegawa98,tonegawa01}. 
In this half-magnetization-plateau region, furthermore, the wave function of the lowest energy state was shown to have the periodicity four in units of the lattice constant, although the periodicity of the original Hamiltonian is two~\cite{tonegawa98,tonegawa01,totsuka98}. The result indicates that spontaneous translational-symmetry breaking occurs~\cite{tonegawa98,tonegawa01,totsuka98}. This feature is consistent with the necessary condition for the appearance of the half-magnetization-plateau in the $S=1/2$ spin chain \cite{oshikawa97}. 
The stability of the half-magnetization-plateau was discussed using bosonization technique combined with the renormalization-group theory and the perturbation calculation~\cite{totsuka98}. On the basis of the results, the phase diagram of the half-magnetization-plateau, which is consistent with the numerical results~\cite{tonegawa98,tonegawa01}, was obtained as a function of the parameters of the bond-alternation and the NNN interaction~\cite{totsuka98}.
A simple picture of the half-magnetization-plateau state was presented, indicating the two-fold degenerate states with singlet and triplet pairs occupying strong bonds alternately~\cite{totsuka98}. 
Furthermore, it was shown that the elementary excitation is given by a superposition of massive kinks and anti-kinks which have $S^z=\pm 1/2$, respectively~\cite{totsuka98}.

The dynamical structure factor (DSF) provides us with important information about the intensity of the magnetic excitation as a function of energy-momentum transfer. Since the DSF can be observed experimentally by inelastic neutron-scattering measurements, a detailed comparison between theoretical and experimental results has been developed. 
The DSF of the $S=1/2$ bond-alternating spin chain with a NNN interaction in magnetic fields was investigated~\cite{Poil,MH,YH00} in connection with experimental findings of the spin-Peierls material ${\rm CuGeO_3}$ ~\cite{Fagot}, since this model with adequate parameters can be an effective model for ${\rm CuGeO_3}$~\cite{RD,CCE}. 
Critical and dynamical properties of typical (quasi-)1D $S=1/2$ spin-gapped systems such as the bond-alternating spin chain, the spin chain with a NNN interaction, and the two-leg ladder system were investigated systematically on the basis of bosonization technique~\cite{giamarchi97}. 
It was shown that the DSF exhibits different peak structures among these three systems, when the gap is collapsed by magnetic fields. 
The DSF of the $S=1/2$ spin chain with a NNN interaction was calculated at half of the saturated magnetization using the numerical diagonalization method~\cite{Wup2}. The distribution of the intensity was discussed, putting stress on the behavior around the field-induced soft mode. 
One of the present authors calculated the DSF of the half-magnetization-plateau state in the $S=1/2$ bond-alternating spin chain with a NNN interaction using two sets of parameters~\cite{usami00}. It was shown that the periodicity and the amplitude of the dispersion relation of the transverse DSF are sensitive to the parameters. 
In spite of this investigation, the dynamical property of the half-magnetization-plateau state has not been fully understood. A detailed investigation is desirable. 

In this paper, we investigate dynamical properties of the half-magnetization-plateau state in the $S=1/2$ bond-alternating spin chain with a NNN interaction. 
According to the phase diagram~\cite{tonegawa98,tonegawa01,totsuka98}, we calculate the DSF systematically in the parameter space of the half-magnetization-plateau. In \S2, we briefly summarize the method for the numerical calculation. In \S3, we show the results for the transverse DSF. In the half-magnetization-plateau state with a strong bond-alternation and a weak NNN interaction, the Hamiltonian of the system can be mapped onto an effective Hamiltonian described by the 1D $S=1/2$ Heisenberg-Ising model in zero magnetic field~\cite{totsuka98,mila98}. We compare the numerical results with the exact excitation spectrums of the effective Hamiltonian. We also estimate the finite-size effects carefully to investigate characteristics of the lowest excited states. We further investigate the static structure factor (SSF) in the half-magnetization-plateau state. Section IV is devoted to the summary of the paper. 
%
%
%
%
%
\section{Model and Method}
We consider the $S=1/2$ bond-alternating spin chain with a NNN interaction in magnetic fields described by the following Hamiltonian, 
%
\begin{eqnarray}
{\cal H} &=& {\cal H}_{0} + {\cal H}_{Z} , 
\label{eqn:s2-1} \\
{\cal H}_{0} &=& J\sum_{i=1}^{N} \left\{ [1-(-1)^{i}\delta] \mbox{\boldmath$S$}_{i} \cdot \mbox{\boldmath$S$}_{i+1} + \alpha \mbox{\boldmath$S$}_{i} \cdot \mbox{\boldmath$S$}_{i+2} \right\} , 
\label{eqn:s2-2} \\
{\cal H}_{Z} &=& -g\mu_B H \sum_{i=1}^{N} S_{i}^{z} , 
\label{eqn:s2-3}
\end{eqnarray}
%
\noindent
where $\delta$ denotes the bond-alternation, $\alpha$ denotes the NNN interaction, $N$ is the total number of the site, and $H$ is magnetic field. 
We set $J=1$ and $g\mu_B=1$. The periodic boundary condition is applied. 

In magnetic field along the $z$ axis, rotational symmetry around the $x$ and $y$ axes is broken, while that around the $z$ axis remains. Therefore, the Hamiltonian can be classified into the subspace according to the magnetization $m=M/N$ with $M=\sum_{i=1}^{N} S_{i}^{z}$. We set that $|\Psi_{0,m}\rangle$ is the eigenfunction of the lowest eigenvalue of ${\cal H}$ with magnetization $m$. 
In given $m$, the DSF can be expressed in the form of the continued fraction \cite{gagliano87} as 
%
\begin{eqnarray}
S^{\mu}(q,\omega) 
   &=& \langle \Psi_{0,m}| S_{q}^{\mu \dagger} 
       \frac{1}{z^{\mu}-{\cal H}} S_{q}^{\mu} |\Psi_{0,m} \rangle 
                \nonumber \\
   &=&  S^{\mu}(q) C^{\mu}(q,\omega), 
   \hspace{8mm}(\mu = +, -, z) 
\label{eqn:s2-4} \\
S^{\mu}(q) &=& \langle\Psi_{0,m}| S_{q}^{\mu \dagger} 
                                S_{q}^{\mu} |\Psi_{0,m}\rangle, 
\label{eqn:s2-5} 
\end{eqnarray}
%
\begin{eqnarray}
\lefteqn{C^{\mu}(q,\omega)} \nonumber \\ 
  &=& 
-\frac{1}{\pi} \lim_{\varepsilon \rightarrow +0} 
\frac{1}
{
z^{\mu}-\alpha_{0}
-\frac{\displaystyle \beta_{1}^{2}}{\displaystyle z^{\mu}-\alpha_{1}
-\frac{\displaystyle \beta_{2}^{2}}{\displaystyle z^{\mu}-\alpha_{2}
-\frac{\displaystyle \beta_{3}^{2}}{\displaystyle z^{\mu}-\alpha_{3}
-\cdots
}}}} , \nonumber \\
&& \makebox{}
\label{eqn:s2-6}
\end{eqnarray}
%
\noindent
where $S^{\mu}_{q} = (1/\sqrt{N}) \sum_{j} {\rm e}^{i q j} S^{\mu}_{j}$, $z^{z}=\omega+i\varepsilon+E^{0}_m$, $z^{\pm}=\omega+i\varepsilon+E^{0}_m\pm H$, $\hbar=1$, and $E^{0}_{m}$ is the lowest eigenvalue of ${\cal H}_0$ with magnetization $m$. In Eq. (\ref{eqn:s2-6}), $\alpha_{i}$ and $\beta_{i+1}$ $(i=0,1,2,\cdots)$ are the diagonal and the sub-diagonal elements of the tridiagonalized Hamiltonian ${\cal H}_0$. We put the lattice constant to unity.  
The DSF $S^{x}(q,\omega)$ is obtained by the relation $S^{x}(q,\omega)=\left[ S^{+}(q,\omega)+S^{-}(q,\omega)\right]/4$, where $S^{x}(q,\omega)=S^{y}(q,\omega)\ne S^{z}(q,\omega)$. 
We can determine poles and residues of this continued fraction numerically. The total contribution of $C^{\mu}(q,\omega)$ for fixed $q$ is normalized to unity, because the following sum rule has to be satisfied, 
%
\begin{equation}
S^{\mu}(q) = \int_{0}^{\infty} {\rm d}\omega S^{\mu}(q, \omega). 
\label{eqn:s2-8}
\end{equation}
%

To calculate $S^{\mu}(q,\omega)$ according to (\ref{eqn:s2-4}), we have to set the magnetic field $H_P$ of the half-magnetization-plateau region. 
We now express $E^{0}_m$ as $E^0(N,M)$ in the $N$-spin system, fixing $m=1/4$. 
We define the magnetic field where the magnetization changes from $M$ to $M \pm 1$ as~\cite{comm} 
%
\begin{eqnarray}
H^{+} = E^{0}(N,M+1)-E^{0}(N,M) ,
\label{eqn:s2-10} \\
H^{-} = E^{0}(N,M)-E^{0}(N,M-1) ,
\label{eqn:s2-11}
\end{eqnarray}
%
where $H^{+}$ and $H^{-}$ are the magnetic fields at the highest and the lowest ends of the magnetization plateau, respectively~\cite{sakai91,tonegawa98}. 
Using $H^{+}$ and $H^{-}$, we can express the magnetic field $H_P$. 
Here, we set $H_{P} = [H^{+}+H^{-}]/2$.

We first calculate $|\Psi_{0,m}\rangle$, $E^{0}(N,M)$, and $E^{0}(N,M\pm 1)$ numerically with $M=N/4$, fixing $N$. The Hamiltonian is then tridiagonalized by the Lanczos method. We substitute $H_{P}$, $\alpha_{i}$ and $\beta_{i+1}$ for the expression (6). Instead of taking the limitation $\varepsilon\rightarrow +0$, we set $\varepsilon=3.0\times10^{-2}$. The convergency of the continued fraction can be estimated quantitatively by the modified Lenz method~\cite{lenz}. 
The finite-size effects are estimated in the same way as that used in Refs. 18 and 19.  
In this way, we obtain the DSF in the half-magnetization-plateau state.

\section{Numerical Results}

\subsection{DSF in the half-magnetization-plateau state}
In Fig. 1, the transverse DSF $S^{x}(q,\omega)$ in the half-magnetization-plateau state is shown for N=28 and $M=7$. The intensity of $S^{x}(q,\omega)$ is represented using the same scale, and is proportional to the area of the full circle. The accuracy of the continued fraction in (6) becomes worse, when we treat the higher energy region. To get accurate results, we have increased the order of the continued fraction. 
As a result, the continued fraction has converged within the relative error $O(10^{-10})$ for $\omega <1$, and $O(10^{-2})$ for $\omega \geq3$. 

 In larger $\delta$ and/or smaller $\alpha$, the separation of the excitation bands becomes apparent, and we can see nearly three excitation bands in the region $\omega <1$, $\omega \sim 2$, and $\omega \geq3$. 
The width of the excitation band strongly depends on the parameters $\alpha$ and $\delta$. Furthermore, narrow and almost dispersionless bands appear in some adequate sets of $\alpha$ and $\delta$ close to $2\alpha+\delta=1$. Concerning this feature, we will show details later. 
The larger intensity lies mainly in the lowest excitation bands. The wavenumber of the largest intensity in given $\alpha$ shifts from $\pi$ to $\pi/2$, as $\delta$ increases. The results indicate that the effect of the bond-alternation becomes dominant. 
Thus, we turn our attention to the behavior of the lowest excitation band.  

We have also calculated the longitudinal DSF $S^{z}(q,\omega)$, and have found that $S^{z}(q,\omega)$ does not exhibit noticeable features depending on the parameters $\alpha$ and $\delta$. Therefore, we consider only the behavior of $S^{x}(q,\omega)$ in the following.

\subsection{Comparison with the excitation spectrums of the effective Hamiltonian}
We first compare the distribution of the intensity of $S^x(q, \omega)$ in the lowest excitation band with the exact excitation spectrums for the effective Hamiltonian. 
In the half-magnetization-plateau state with $\delta \sim 1$ and $\alpha \sim 0$, the model described by (\ref{eqn:s2-1}) can be mapped onto the 1D $S=1/2$ Heisenberg-Ising model in zero field described as~\cite{totsuka98,mila98}
%
\begin{equation}
\tilde{{\cal H}}=\sum_{j=1}^{N/2} \left\{
\tilde{J}_{xy}(\tilde{S}_{j}^{x} \tilde{S}_{j+1}^{x} 
             + \tilde{S}_{j}^{y} \tilde{S}_{j+1}^{y})
+\tilde{J}_{z} \tilde{S}^{z}_{j} \tilde{S}^{z}_{j+1}
\right\} , 
\label{eqn:s3-1}
\end{equation}
%
\noindent
where $\tilde{J}_{xy}=[2\alpha-(1-\delta)]/2$ and $\tilde{J}_{z}=[2\alpha+(1-\delta)]/4$. 
The effective spin-$1/2$ operators are 
$\tilde{S}_i^z = 2(S_{2l-1}^z-1) = 2(S_{2l}^z-1), 
\tilde{S}_{i}^{+} = \sqrt{2}S_{2l-1}^+ = -\sqrt{2}S_{2l}^+$, and 
$\tilde{S}_{i}^{-} = \sqrt{2}S_{2l-1}^- = -\sqrt{2}S_{2l}^- , \; (l=1,2,3, \cdots, N/2)$. 
The ground state of (\ref{eqn:s3-1}) is a N\'eel-like state with two-fold degeneracy, which corresponds to the singlet and triplet pairs occupying strong $(1+\delta)$-bonds alternately in the original Hamiltonian~\cite{totsuka98}. 
The elementary excitation, which keeps $S^z=\pm1$ and $0$, can be given by a superposition of massive kinks and anti-kinks which have $S^z=\pm1/2$, respectively.

In Fig. 2, the results for N=28 and $M=7$ are shown. 
The solid lines represent the exact bounds of the elementary excitation for the effective Hamiltonian obtained by the Bethe ansatz method~\cite{bougourzi98,takahashi}. 
In smaller $\delta$ of given $\alpha$, the intensity distributes beyond the upper and lower bounds of the excitation continuum for the effective Hamiltonian.  In larger $\delta$ of given $\alpha$, on the other hand, the intensity distributes within the excitation continuum for the effective Hamiltonian. 
In the latter case, the ground state can be given by the singlet and triplet pairs occupying strong $(1+\delta)$-bonds alternately, and the elementary excitation can be described by a superposition of massive two kinks and anti-kinks which have $S^z=\pm1/2$. The schematic picture for the elementary excitation is the same as that shown in Fig. 5 of Ref. 3, where the kink and anti-kink are represented by domain-wall-like excitations from the ordered ground state mentioned above. 
In the former case, it is not easy to present a schematic picture for the elementary excitation. The higher order terms which are released to obtain the effective Hamiltonian may become important. 
It is interesting that even in $\alpha > 0.1$, the intensity distributes within the excitation continuum for the effective Hamiltonian, although the mapping onto the effective Hamiltonian can be successful for $\delta \sim 1$ and $\alpha \sim 0$. 

In the parameter region where $2\alpha+\delta=1$ is satisfied, the effective Hamiltonian is reduced to the 1D $S=1/2$ Ising model, and then the elementary excitation shows a dispersionless mode. Note that this parameter region in zero field is called the Shastry-Sutherland line~\cite{shastry81}. 
In Fig. 2, we can see $S^x(q, \omega)$ in the parameter region $2\alpha+\delta=1$; $(\alpha, \delta) = (0.1, 0.8), (0.2, 0.6), (0.3, 0.4)$ and $(0.4, 0.2)$, where the dispersionless mode of the effective Hamiltonian is represented by the broken line in given $\alpha$.  
At $\delta=0.8$ and $\alpha=0.1$, the agreement between the numerical results and the results from the effective Hamiltonian is excellent, indicating that the lowest excitation is described by a superposition of massive two kinks and anti-kinks, and is well localized in the real space. 
As the parameter region is removed from the successful mapping point where $\delta \sim 1$ and $\alpha \sim 0$, the intensity spreads more widely and the spatial localization of the elementary excitation becomes worse.

To see the characteristics of the low-lying excitations in detail, we next investigate the finite-size effects of the lowest and the next-lowest excited states. According to Refs. 18 and 19, we investigate the size dependence of the poles and their residues of the continued fraction $C^{x}(q,\omega)$, to discuss whether a pole belongs to an excitation continuum or forms an isolated mode. 
When a pole belongs to an excitation continuum, at least either of its position or its residue has appreciable size dependence. On the other hand, when a pole forms an isolated mode, the position and the residue of a pole has little size dependence. 
This method was successfully used to investigate the isolated mode of the $S=1$ Haldane gap system~\cite{taka}. Later on, characteristics of the lowest excited state for the $S=1/2$ bond-alternating spin chain with a NNN interaction was investigated using the same method~\cite{ys}. In this study, it was shown that a definite conclusion concerning the isolated mode is difficult to draw only from this procedure. 

In Fig. 3, we show the typical results for the system-size dependence of poles and residues of the continued fraction in the present system. 
Note that we have confirmed that the same system-size effects already obtained for the poles and residues of the $S=1/2$ $XXX$ chain~\cite{taka,ys} can be reproduced. 

We first consider the behaviors in the region where the intensity spreads within the excitation continuum of the effective Hamiltonian. The typical results are shown in Fig. 3 (a), which corresponds to the bottom figure in Fig. 2 (c).  In this region, the size dependence of the residues is not noticeable. The positions of the poles for the lowest excited states show almost no size dependence. Yet, the positions of the poles for the next lowest states decrease, as the system size increases. 
Judging from these results and the fact that the intensity spreads within the excitation continuum, the lowest mode of the present region is the lower bound of the excitation continuum. 

We next consider the behaviors in the region where the intensity spreads beyond the excitation continuum. In this region, it is difficult to present a definite conclusion whether the lowest excited states form an isolated mode. 
As shown in Figs. 3 (b) and (c), which correspond to the top figures in Figs. 2 (b) and (d) respectively, we cannot see clear size dependence of the residues. The positions of the poles for the lowest excited states scarcely show size dependence. Nevertheless, the positions of the poles for the next lowest states decrease around $q=0, \pi/2$, and $\pi$, or around two wavenumbers among them, as the system size increases. 
Thus, in this parameter region, the lowest excited states may become a lower bound of the excitation continuum around $q=0, \pi/2$, and $\pi$, or around two wavenumbers among them, while they may form an isolated mode in the other wavenumbers.  

In the above arguments, we leave the results for the parameter region where the almost dispersionless mode appears. Analysis in such parameter regions will be performed in the next subsection.

\subsection{DSF in the dispersionless mode}
In the parameter region away from $\delta \sim 1$ and $\alpha \sim 0$, the almost dispersionless mode of $S^{x}(q,\omega)$ appears around $\delta \sim 0.64$ in $\alpha = 0.2$, $\delta \sim 0.5$ in $\alpha = 0.3$, and $\delta \sim 0.38$ in $\alpha = 0.4$ as shown in Fig. 4. 
The size dependence of poles and residues at these points is shown in Fig. 5. 
Note that the similar behaviors are seen between the cases of $(\alpha, \delta) = (0.1, 0.8)$ and $(0.2, 0.64)$, and between $(\alpha, \delta) = (0.3, 0.5)$ and $(0.4, 0.38)$, respectively. 
The residues in $(\alpha, \delta) = (0.1, 0.8)$ and $(0.2, 0.64)$ show no size dependence, while those in $(\alpha, \delta) = (0.3, 0.5)$ and $(0.4, 0.38)$ decrease around $q \sim 0.75\pi$ and $\pi$ with increasing the system size. The positions of the poles in $(\alpha, \delta) = (0.1, 0.8)$ and $(0.2, 0.64)$ show no size dependence. 
Therefore, in $(\alpha, \delta) = (0.1, 0.8)$ and $(0.2, 0.64)$, the lowest excitation is localized spatially and forms apparently the isolated mode below the excitation continuum. 
In fact, no intensity distributes between the lowest dispersionless mode and the excitation band around $\omega \sim 2.0$ in the figure of $(\alpha, \delta) = (0.1, 0.8)$ in Fig. 1 (a). Note that the same occurrence is seen also in $(\alpha, \delta) = (0.2, 0.64)$, although we have not presented the results in Fig. 1 (b). 
On the contrary, in $(\alpha, \delta) = (0.3, 0.5)$ and $(0.4, 0.38)$, the positions of the poles for the lowest excited states show no size dependence, while those for the next lowest states decrease around $q \sim 0.75\pi$ and $\pi$ with increasing the system size. 
Judging from the results, the lowest excitation in $(\alpha, \delta) = (0.3, 0.5)$ and $(0.4, 0.38)$ is probably the lower edge of the excitation continuum around $q \sim 0.75\pi$ and $\pi$, and is the isolated mode in the other wavenumbers. 

In this study, we have neither found the completely isolated mode in $\alpha=0.3$ nor $0.4$. Nevertheless, there may appear the completely isolated mode close to the parameters $(\alpha, \delta) = (0.3, 0.5)$ and $(0.4, 0.38)$. We leave this issue in the future. 

In the previous and present subsections, we have discussed the isolated mode between the ground state and the excitation continuum. The isolated mode can be seen in several 1D spin-gapped systems such as the Haldane-gap system~\cite{halex,taka,ZLP01}, the $S=1/2$ bond-alternating spin chain with a NNN interaction~\cite{us,spex,ys,MH,koto1}, the $S=1/2$ two-leg spin ladder system~\cite{koto2,jb,legex,uhrig}, and so on~\cite{sach,moni}. In some systems, the isolated mode was observed experimentally~\cite{halex,spex,legex}. It may be interesting that such an isolated mode has appeared also in the field-induced ordered state of the 1D quantum spin system.

\subsection{SSF in the half-magnetization-plateau state}
We calculate the SSF $S^x(q)$ in the half-magnetization-plateau state using the sum rule (\ref{eqn:s2-8}). The results are shown in Fig. 6. The finite-size effects appear only at $q=\pi$ or $q=\pi/2$ where the intensity takes the maximum. 
The largest intensity is located at $q=\pi$ in the small $\delta$ region.  As $\delta$ increases, the wavenumber of the largest intensity shifts towards $q=\pi/2$, taking the incommensurate value of $q$. This occurrence reflects that the bond-alternation becomes effective.  
In $\alpha=0.4$, the region of the incommensurate wavenumber becomes wider as compared to that for other $\alpha$'s. 

In other 1D spin-gapped systems such as the Haldane-gap system~\cite{taka} and the $S=1/2$ Heisenberg chain with a NNN interaction~\cite{ys}, the incommensurate mode between $\pi/2$ and $\pi$ is difficult to observe. 
The dominant incommensurate mode of the transverse spin correlation may be characteristic of the magnetization-plateau state in the present system.

\section{Summary}
%
We have investigate the DSF of the half-magnetization-plateau state in the $S=1/2$ bond-alternating spin chain with a NNN interaction. 
The results are summarized in the phase diagram of Fig. 7. The half-magnetization-plateau appears along the thick lines in given $\alpha$~ \cite{tonegawa98}.  At the full circles,  the intensity distributes within the excitation continuum for the effective Hamiltonian, while at the open circles, the intensity distributes beyond the excitation continuum. 
In the former regions, the ground state and the elementary excitation are probably described by a simple picture based on the effective Hamiltonian. 
From the analysis of the size dependence, we have concluded that the lowest excited states in the former region are the lower bound of the excitation continuum, while those in the latter region may form the isolated mode below the excitation continuum. 
Along the dotted line, the dispersionless behavior comes out, indicating that the elementary excitation is localized spatially. From the size dependence, we have concluded that the dispersionless mode is isolated below the excitation continuum. 

Above the upper solid line, the largest intensity of SSF is located  at $q=\pi/2$, and below the lower solid line, the largest intensity of SSF is located at $q=\pi$. As $\delta$ increases between two solid lines, the wavenumber of the largest intensity shifts from $q=\pi$ to $q=\pi/2$, taking the incommensurate value. The results indicate that with increasing $\delta$ in fixed $\alpha$, the dominant mode of the transverse spin correlation shifts $\pi$ to $\pi/2$, taking an incommensurate value.

%
%
%

\section{acknowledgments}

We would like to thank I. Harada, M. Takahashi and T. Tonegawa for useful comments and valuable discussions. 
Our computational programs are based on TITPACK Ver. 2 by H. Nishimori. 
Numerical computation was partly carried out at the Yukawa Institute Computer Facility, Kyoto University, and the Supercomputer Center, the Institute for Solid State Physics, University of Tokyo. 


\begin{references}
\bibitem{comm}
To calculate the precise values of the magnetic fields in the thermodynamic limit at both ends of the magnetization plateau, finite-size-scaling analysis based on conformal field theory is worthwhile as shown in Refs. \cite{sakai91,tonegawa98}. Nevertheless, we have neglected the finite-size correction in (\ref{eqn:s2-10}) and (\ref{eqn:s2-11}), because the error in the magnetic fields thus obtained shifts the DSF slightly  along the $\omega$ axis.
\end{references}

%
%
%

%
%
%
%
%

\end{document}